\documentclass[aps,amsmath,notitlepage,twocolumn,amssymb,prl,longbibliography]{revtex4-1}
\usepackage{graphicx}
\usepackage{dcolumn}
\usepackage{bm}
\usepackage[colorlinks=true,linkcolor = blue,citecolor=blue,urlcolor=blue,anchorcolor = blue]{hyperref}
\usepackage{wasysym}
\usepackage{stmaryrd}
\usepackage{verbatim}
\usepackage{subfigure}
\usepackage{amsmath}
\usepackage{times}
\usepackage[version=4]{mhchem}
\usepackage{braket}
\usepackage{siunitx}
\usepackage{amssymb}
\usepackage{color}
\usepackage{multirow}
\usepackage{tabularx}
\usepackage{amsthm,amsmath,amssymb}
\usepackage{mathrsfs}
\usepackage{hyperref}

\newcommand{\RNum}[1]{\uppercase\expandafter{\romannumeral #1\relax}}

\newcolumntype{Y}{>{\centering\arraybackslash\hsize=.6\hsize}X}
\newcolumntype{Z}{>{\centering\arraybackslash\hsize=.13\hsize}X}

\begin{document}

\title{Crystalline electric field excitations and their nonlinear splitting under magnetic fields in \ce{YbOCl}}

\author{Yanzhen\,Cai$^{1,2}$}
\author{Wei\,Ren$^{1,2}$}
\author{Xijing\,Dai$^{1}$}
\author{Jing\,Kang$^{2}$}
\author{Weizhen\,Zhuo$^{1,2}$}
\author{Mingtai\,Xie$^{1,2}$}
\author{Anmin\,Zhang$^{1}$}
\author{Jianting\,Ji$^{2}$}
\author{Feng\,Jin$^{2}$}
\author{Zheng\,Zhang$^{2}$}
\email{zhangzheng@iphy.ac.cn}
\author{Qingming\,Zhang$^{1,2}$}
\email{qmzhang@iphy.ac.cn}

\affiliation{$^{1}$School of Physical Science and Technology, Lanzhou University, Lanzhou 730000, China}
\affiliation{$^{2}$Beijing National Laboratory for Condensed Matter Physics, Institute of Physics, Chinese Academy of Sciences, Beijing 100190, China}

\begin{abstract}
Recently reported van der Waals layered honeycomb rare-earth chalcohalides \ce{REChX} (RE = rare earth, Ch = chalcogen, and X = halogen) are considered to be promising Kitaev spin liquid (KSL) candidates. The high-quality single crystals of \ce{YbOCl}, a representative member of the family with an effective spin of 1/2, are available now. The crystalline electric field (CEF) excitations in a rare-earth spin system are fundamentally important for understanding both finite-temperature and ground-state magnetism, but remain unexplored in \ce{YbOCl} so far. In this paper, we conduct a comprehensive Raman scattering study to unambiguously identify the CEF excitations in \ce{YbOCl} and determine the CEF parameters and wave functions. Our Raman experiments further reveal the anomalous nonlinear CEF splitting under magnetic fields. We have grown single crystals of \ce{YbOCl}, the nonmagnetic \ce{LuOCl}, and the diluted magnetic \ce{Lu_{0.86}Yb_{0.14}OCl} to make a completely comparative investigation. Polarized Raman spectra on the samples at 1.8 K allow us to clearly assign all the Raman-active phonon modes and explicitly identify the CEF excitations in \ce{YbOCl}. The CEF excitations are further examined using temperature-dependent Raman measurements and careful symmetry analysis based on Raman tensors related to CEF excitations. By applying the CEF Hamiltonian to the experimentally determined CEF excitations, we extract the CEF parameters and eventually determine the CEF wave functions. 
The study experimentally pins down the CEF excitations in the Kitaev compound \ce{YbOCl} and sets a foundation for understanding its finite-temperature magnetism and exploring the possible nontrivial spin ground state.

\end{abstract}

\maketitle

\subsection{Introduction}

By constructing bond-dependent anisotropic Ising-like spin-exchange interactions on the honeycomb lattice, an exactly solvable Kitaev model can be obtained \cite{2}.
The model represents a groundbreaking framework in the study of quantum spin liquids (QSL) and topological quantum computations.
One of the most intriguing aspects of the Kitaev model is its potential to host non-Abelian anyons \cite{PhysRevLett.129.037201,PhysRevLett.132.206501,PhysRevB.109.125150}, which are a type of quasiparticle that can be used for fault-tolerant quantum computations \cite{PhysRevA.109.032421,10.1063/5.0102999}. 
Therefore the search for Kitaev spin liquid (KSL) candidate materials has attracted significant interest.
Rare-earth ions exhibit high magnetic anisotropy originating from strong spin-orbit coupling (SOC).
The Jackeli-Khaliullin mechanism inspires us to search for KSL candidate materials in rare-earth compounds \cite{PhysRevLett.102.017205,PhysRevLett.110.097204}.

Recent investigations have identified the rare-earth chalcohalides REChX family as promising candidates for KSL materials. These compounds have garnered significant interest due to their distinctive structural and magnetic properties, positioning them as ideal systems for probing KSL states \cite{JiantingJi,PhysRevResearch.4.033006}.
Most of these compounds exhibit high symmetry characterized by the \(R\bar{3}m\) space group, with the nearest-neighbor rare-earth magnetic ions forming a perfectly undistorted honeycomb lattice.
\ce{YbOCl} is a prototypical material within the family \cite{PhysRevResearch.4.033006}. 
Particularly, \ce{Yb^{3+}} ions having an odd number of 4$f$ electrons possess a doubly degenerate CEF ground state (Kramers doublets), which is protected by time-reversal symmetry and yields an effective spin-1/2 required by KSL.
Furthermore, the magnetic ion layers in \ce{YbOCl} are stacked through van der Waals interactions.
This type of van der Waals structure exhibits excellent two-dimensional properties, enabling the realization of Kitaev physics in few-layer or even single-layer honeycomb lattices \cite{PhysRevResearch.4.033006,PhysRevB.104.214410}. 

For this purpose, we have successfully grown high-quality single crystals of \ce{YbOCl}, with a maximum size of approximately 15 mm. This allows us to comprehensively understand the magnetism of \ce{YbOCl} in different aspects.
In the previous work, we conducted a comprehensive study on the magnetism of \ce{YbOCl} at finite temperatures and in its ground state\cite{PhysRevResearch.4.033006}.
Experimental results indicate that \ce{YbOCl} undergoes a magnetic phase transition at 1.3 K, with its ground state exhibiting A-type antiferromagnetism (AFM).
More interestingly, a magnetic field of approximately 0.3 T along the \(c\) axis can induce \ce{YbOCl} into a spin-disordered state \cite{YbOCl}.
Although we have gained a comprehensive understanding of the ground state magnetism of \ce{YbOCl}, there is still a lack of quantitative measurements and studies on its CEF excitations.
For rare-earth magnetic ions, the CEF plays a crucial role in magnetism.
Therefore it is necessary to conduct a comprehensive study of CEF excitations in \ce{YbOCl}.

In this paper, the CEF excitations of \ce{YbOCl} are comprehensively measured and analyzed using Raman scattering technique.
First, we performed temperature-dependent x-ray diffraction (XRD) measurements on \ce{YbOCl}. Through structural refinement of these XRD patterns, we ruled out the presence of temperature-induced structural phase transitions in \ce{YbOCl}.
This exclusion allowed us to focus solely on the CEF excitations of \ce{YbOCl} without the complicating factors of structural changes.

Subsequently, we performed Raman scattering measurements on \ce{YbOCl}, the nonmagnetic control sample \ce{LuOCl}, and the diluted magnetic sample \ce{Lu_{0.86}Yb_{0.14}OCl} at 1.8 K.
By combining symmetry analysis and Raman scattering tensors of phonons, we identified all phonon excitation peaks in these three materials.
More importantly, we also observed three additional excitation peaks in the Raman scattering spectrum of \ce{YbOCl} near 319, 327, and 523 cm$^{-1}$, which we preliminarily identified as the CEF excitations of \ce{YbOCl}.
To confirm more reliably that these three additional peaks are CEF excitations of \ce{YbOCl}, we provided further evidence from multiple perspectives.

\begin{figure*}[t]
	\includegraphics[scale=0.5]{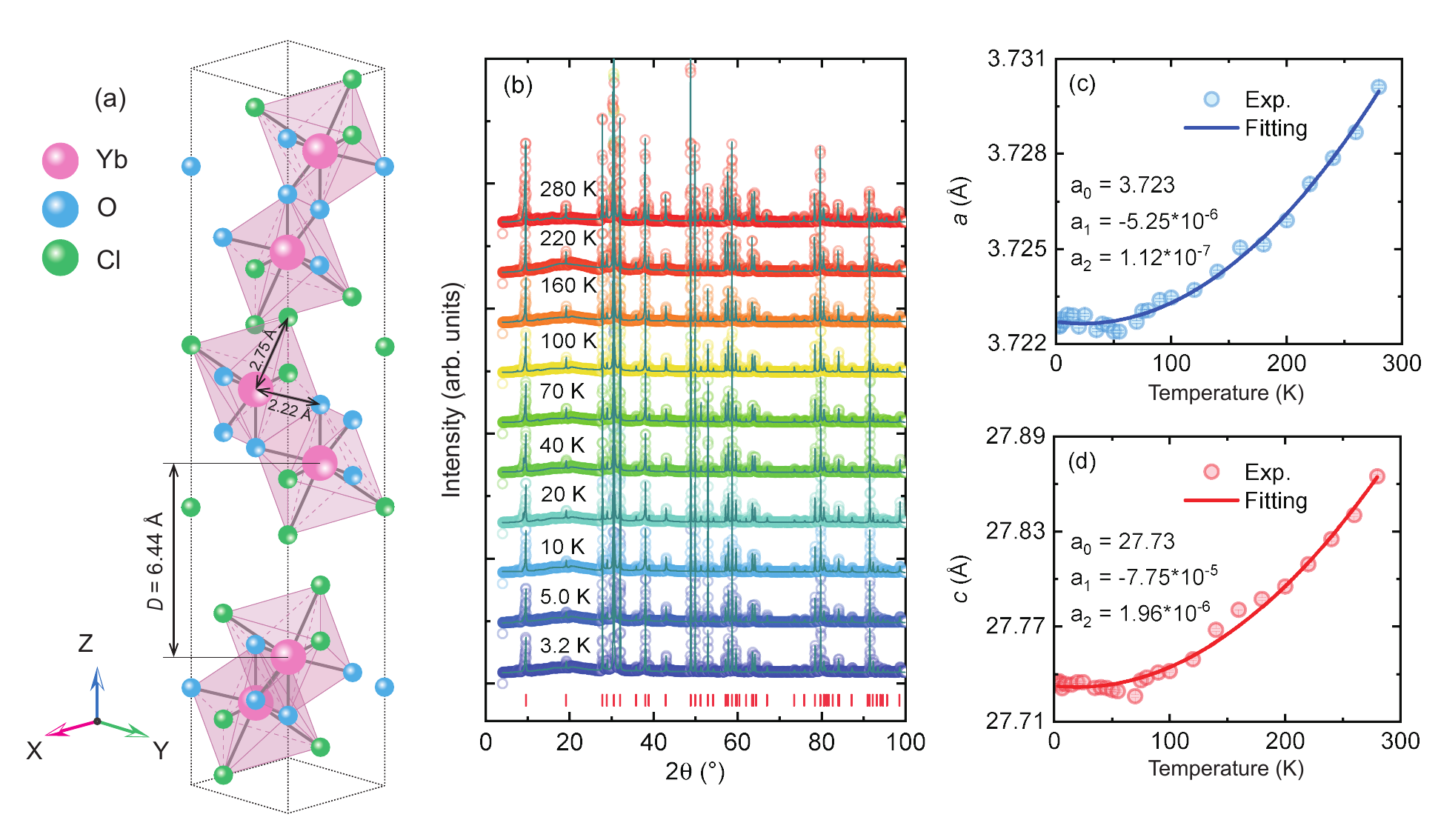}
	\caption{\label{fig:epsart}{Schematic diagram of the \ce{YbOCl} crystal structure and temperature-dependent XRD measurements. (a) The crystal structure of \ce{YbOCl}. (b) The temperature-dependent XRD patterns of \ce{YbOCl}. The open circles represent the experimental measurement results. The green solid lines represent the refined XRD results. The temperature-dependent lattice parameters $a$ (c) and $c$ (d) were obtained from refined XRD. The open circles represent data from XRD refinement, and the solid lines are the fitting results from the empirical formula  \hyperref[for:1]{(1)}}.}
	\label{fig:Fig1}
\end{figure*}

From a basic understanding, \ce{YbOCl} is a good insulator, and the phase transition temperature of \ce{YbOCl} is 1.3 K. Therefore within the range of several hundred wavenumbers, the excitation peaks, apart from phonon peaks, are only CEF excitations.
Besides, temperature-dependent Raman scattering experiments on \ce{YbOCl} did not show significant shifts in these three excitation peaks over temperature which is consistent with the temperature dependence characteristics of CEF excitations.
It should be noted that although \ce{YbOCl} is a good van der Waals material, the effect of van der Waals interactions on the CEF can be practically ruled out.
The primary reason is that van der Waals interactions are higher-order dipole interactions, whereas the CEF primarily originates from the electrostatic potential. The energy scales of the two are significantly different. Moreover, the bond lengths of Yb-O and Yb-Cl, which are 2.22 Å and 2.75 Å respectively, are shorter than the interlayer distance $D$ = 6.44 Å [see Fig. \hyperref[fig:Fig1]{1(a)}].
We also calculated the CEF excitation levels of \ce{YbOCl} from the perspective of the point charge model (PCM). The calculated CEF first and second excitation energy levels are close to the results obtained from the experimental measurements.
In \ce{Lu_{0.86}Yb_{0.14}OCl}, weak excitations were also observed near the two lowest additional excitation peaks of \ce{YbOCl}. The intensity of these excitations is approximately ten times smaller than that of \ce{YbOCl}.
This also serves as indirect evidence supporting the CEF excitations in \ce{YbOCl}.
The most important and compelling evidence is our analysis based on the symmetry of CEF wave functions and CEF Raman tensors.
The intensity of polarized Raman scattering under different polarizations is completely consistent with our analysis through the CEF Raman tensors.

We also investigated the CEF excitations of \ce{YbOCl} along the \(c\) axis and in the \(ab\) plane under different magnetic fields.
By applying external magnetic fields, we observed an obvious nonlinear splitting effect in the CEF first and second excitation energy levels.
Incorporating the contributions of internal magnetic fields through the mean-field (MF) approximation, we were able to better explain the nonlinear splitting of these energy levels by the influence of an external magnetic field.

More generally, it is interesting to make a comparison between Raman scattering techniques and inelastic neutron scattering (INS) in studying CEF excitations. Both are inelastic scattering techniques sharing formally similar scattering cross sections.
INS is undoubtedly the most common method for studying CEF excitations in rare-earth materials. It can directly give the momentum-resolved CEF levels.
On the other hand, INS requires a larger quantity of samples to obtain a better signal-to-noise ratio, while even micron-sized samples can be detected with clear excitation signals by Raman scattering. Moreover, Raman scattering offers a much higher energy resolution, which makes it possible to study the CEF splitting under magnetic fields and the coupling between CEF excitations and phonons.
\subsection{\textbf{Samples, experimental techniques, and numerical methods}}

High-quality single crystal samples of \ce{YbOCl}, \ce{LuOCl}, and \ce{Lu_{0.86}Yb_{0.14}OCl} were synthesized through the high-temperature flux method. 
The typical size of these single crystal samples is 4 $\times$ 4 $\times$ 0.05 mm$^3$.
The synthesis method for single crystals of \ce{YbOCl} is as follows.
Anhydrous \ce{YbCl3} and \ce{Yb2O3} were mixed in a mass ratio of 1:4, transferred to a graphite crucible, and then encapsulated in a vacuum quartz tube.
The encapsulated sample was heated to 1050 °C in a muffle furnace and held for 72 hours. Then it was cooled at a rate of 6 °C per hour until reaching the melting point of \ce{YbCl3} (875 °C) \cite{Beck+1976+1562+1564,BRANDT1974411}, and finally cooled naturally to room temperature.
The synthesis method for single crystals of \ce{LuOCl} and \ce{Lu_{0.86}Yb_{0.14}OCl} is similar to that of \ce{YbOCl}, cooling at a rate of 6 °C per hour until reaching the melting point of \ce{LuCl3} (905 °C).
The synthesized sample was treated with deionized water and dilute hydrochloric acid to remove surface impurities, ultimately yielding a transparent single crystal.
These single crystal samples of \ce{YbOCl}, \ce{LuOCl}, and \ce{Lu_{0.86}Yb_{0.14}OCl} were used for Raman scattering experiments.

We also synthesized polycrystalline samples of \ce{YbOCl} for temperature-dependent XRD measurements.
The synthesis method for polycrystalline samples of \ce{YbOCl} is streamlined. The mixture of \ce{YbCl3} and \ce{Yb2O3} is heated at 1050 °C for 72 hours, followed by cooling to room temperature \cite{doi:10.1021/cm00040a010}, yielding the desired sample.

The temperature-dependent XRD data were collected using a Huber G670 high-resolution fast x-ray powder diffractometer equipped with a Zephyr cryostat.
We measured the elemental composition of \ce{YbOCl}, \ce{LuOCl}, and \ce{Lu_{0.86}Yb_{0.14}OCl} samples using a SU5000 scanning electron microscope equipped with a BRUKER XFlash 6160 energy-dispersive spectrometer.
The Raman spectra were collected using an HR800 Evolution (Jobin Yvon) equipped with 633 nm and 473 nm lasers, charge-coupled device (CCD), and volume Bragg gratings.
After cleavage, the single crystals of \ce{YbOCl}, \ce{LuOCl}, and \ce{Lu_{0.86}Yb_{0.14}OCl} were placed in a closed-cycle cryostat (AttoDRY 2100) equipped with a superconducting magnet up to 9 T. The excitation laser beam was focused into a spot of $\sim$ 5 $\mu$m in diameter in the \(ab\) plane of single crystal samples with a power below 0.5 mW to avoid overheating.
Magnetization measurements along the \(c\) axis and in the \(ab\) plane of single crystals of \ce{YbOCl} were performed by a Quantum Design Physical Property Measurement System (PPMS) equipped with a vibrating sample magnetometer (VSM) at 1.8 K.

Based on the CEF theory and diagonalization techniques, we have developed a customized program package for analyzing the zero-field (ZF) CEF excitation energy levels and the magnetic field-dependent splitting of the CEF energy levels.

\begin{figure*}[t]
	\includegraphics[scale=1]{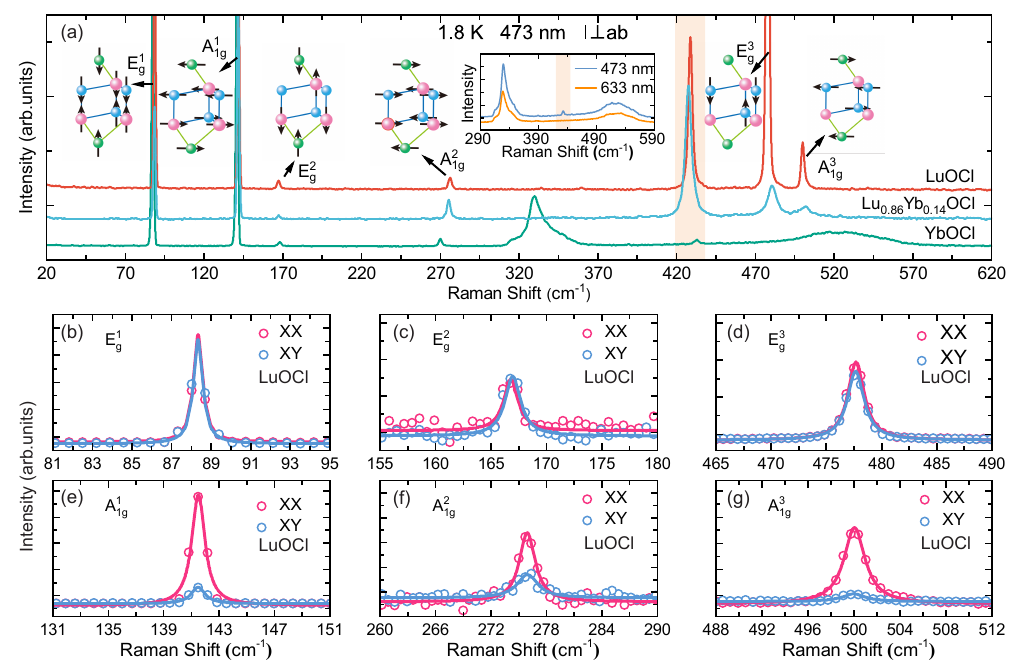}
	\caption{\label{fig:epsart}Comparison of Raman spectra of \ce{YbOCl}, \ce{LuOCl}, and \ce{Lu_{0.86}Yb_{0.14}OCl} at 1.8 K. {(a) The excitation peak in the shaded area is fluorescence excitation, which can be eliminated by selecting different laser wavelengths, as shown in the inset of (a). The vibration patterns in (a) correspond to the vibration modes of different phonon peaks in sequence. [(b)-(g)] The polarized Raman spectra of \ce{LuOCl} at 1.8 K. The XX and XY configurations represent parallel and cross polarization configurations, respectively.}}
	\label{fig:Fig2}
\end{figure*}
\subsection{Crystal Structure, XRD, and Raman active phonons}
The crystal structure of \ce{YbOCl} possesses a space group symmetry of \(R\bar{3}m\).
The schematic diagram of the crystal structure is shown in Fig. \hyperref[fig:Fig1]{1(a)}.
The quasi-two-dimensional plane of magnetic ions \ce{Yb^{3+}}, which is extended from the local structure, satisfies the symmetry requirements of the hexagonal lattice. Therefore it is a candidate material for studying Kitaev spin systems \cite{JiantingJi,PhysRevResearch.4.033006}.
We also synthesized the nonmagnetic sample of \ce{LuOCl} and the diluted magnetic sample \ce{Lu_{0.86}Yb_{0.14}OCl}.
They are isomorphous with \ce{YbOCl} and both possess \(R\bar{3}m\) space group symmetry. Furthermore, their lattice parameters are minimally different from those of \ce{YbOCl} \cite{https://doi.org/10.1002/jrs.1250260108}, rendering them excellent comparative materials for \ce{YbOCl}.
The nonmagnetic sample of \ce{LuOCl} completely eliminates the CEF excitations and the exchange interactions between magnetic ions. 
When we investigate the phonon peaks with Raman activity in \ce{YbOCl}, \ce{LuOCl} can be a good reference material to help us analyze the peak position, symmetry, and vibration mode of the phonon peaks in \ce{YbOCl}.
The diluted magnetic sample \ce{Lu_{0.86}Yb_{0.14}OCl} almost eliminates the exchange interactions between magnetic ions, but the CEF of magnetic ions still remains. As a control sample of \ce{YbOCl}, it can help us analyze the CEF excitations of \ce{YbOCl}.

Before analyzing the CEF excitations of \ce{YbOCl}, we need to rule out the possible structural phase transition of \ce{YbOCl}.
As shown in Fig. \hyperref[fig:Fig1]{1(b)}, we conducted temperature-dependent XRD measurements on polycrystalline samples of \ce{YbOCl}.
Firstly, we did not observe any additional diffraction peaks in the XRD patterns of \ce{YbOCl} at different temperatures.
All the diffraction peaks detected by x-ray match the peak positions given by theoretical calculations [red indicators in Fig. \hyperref[fig:Fig1]{1(b)}].
This indicates that the polycrystalline samples of \ce{YbOCl} we synthesized possess high quality and further confirms that \ce{YbOCl} is single phase.
Secondly, we performed structural refinement on the XRD patterns at different temperatures. Through the refinement, we found that aside from slight changes in the lattice parameters $a$ and $c$, and the crystallographic symmetry of \ce{YbOCl} keeps \(R\bar{3}m\).
The temperature-dependent lattice parameters $a$ and $c$ are shown in Fig. \hyperref[fig:Fig1]{1(c)} and \hyperref[fig:Fig1]{1(d)}.
The lattice parameters $a$ and $c$ did not show any rapid increase or decrease with temperature, indicating that there is no temperature-induced structural phase transition in \ce{YbOCl}.
Moreover, the change in lattice parameters with temperature can be well described by the following empirical formula \cite{woodard1969x}:
\begin{equation}
	L = a_0 + a_1T + a_2T^2,
	\label{for:1}
\end{equation}
where $a_{0}$, $a_{1}$, and $a_{2}$ are fitting parameters, and $T$ represents temperature.
The temperature-dependent lattice parameters can be well simulated by the formula which further indicates that there is no structural phase transition in YbOCl.

Subsequently, we can identify the phonon vibrations of \ce{YbOCl} through Raman scattering.
We conducted Raman scattering measurements on three samples: \ce{YbOCl}, \ce{Lu_{0.86}Yb_{0.14}OCl}, and \ce{LuOCl}. 
These three samples share the $D_{3d}$ point group and \(R\bar{3}m\) space group \cite{JiantingJi,https://doi.org/10.1002/adfm.201903017,SM}.
The crystal symmetry allows for six Raman-active phonon modes: $3E_{g}$ + $3A_{1g}$. 
Symmetry analysis tells us that the $A_{1g}$ mode is visible only in the parallel polarization configuration (XX), while the $E_{g}$ mode can be observed in XX and cross polarization configurations (XY). The two modes can be clearly identified by the polarized Raman scattering spectra.
The Raman spectra of \ce{YbOCl} (green), \ce{Lu_{0.86}Yb_{0.14}OCl} (blue), and \ce{LuOCl} (red) at 1.8 K are presented in Fig. \hyperref[fig:Fig2]{2(a)}.
We can clearly observe six peaks in the Raman spectrum of \ce{LuOCl} at 1.8 K.
Through polarized Raman scattering [see Figs. \hyperref[fig:Fig2]{2(b)}-\hyperref[fig:Fig2]{2(g)}], the vibration modes corresponding to the six phonon peaks can be distinguished one by one \cite{SM}. 
Moreover, the corresponding vibration patterns are also displayed in Fig. \hyperref[fig:Fig2]{2(a)}.
It should be noted that the Raman peak near 430 cm$^{-1}$ is caused by fluorescence, and this excitation peak can be eliminated by replacing the laser with a different wavelength (see the inset of Fig. \hyperref[fig:Fig2]{2(a)}).
The Raman scattering spectrum of \ce{Lu_{0.86}Yb_{0.14}OCl} is highly similar to that of \ce{LuOCl}.
The only difference is that the scattering intensity of the two vibration modes $E_{g}^{3}$ and $A_{1g}^{3}$ is significantly weakened in \ce{Lu_{0.86}Yb_{0.14}OCl}.
In the Raman scattering spectrum of \ce{YbOCl}, we only observed four phonon peaks.
Moreover, three additional broad excitation peaks were observed at 319 cm$^{-1}$ ($P1$), 327 cm$^{-1}$ ($P2$), and 523 cm$^{-1}$ ($P3$).
Based on our research on the CEF excitations of \ce{NaYbS2} \cite{NaYbS2} and \ce{NaYbSe2} \cite{PhysRevB.103.035144} using Raman scattering techniques, we believe that the three additional excitation peaks are related to the CEF of \ce{YbOCl}.
Next, we will quantitatively analyze these excitation peaks in combination with CEF theory.

\begin{figure*}[t]
	\includegraphics[scale=0.17]{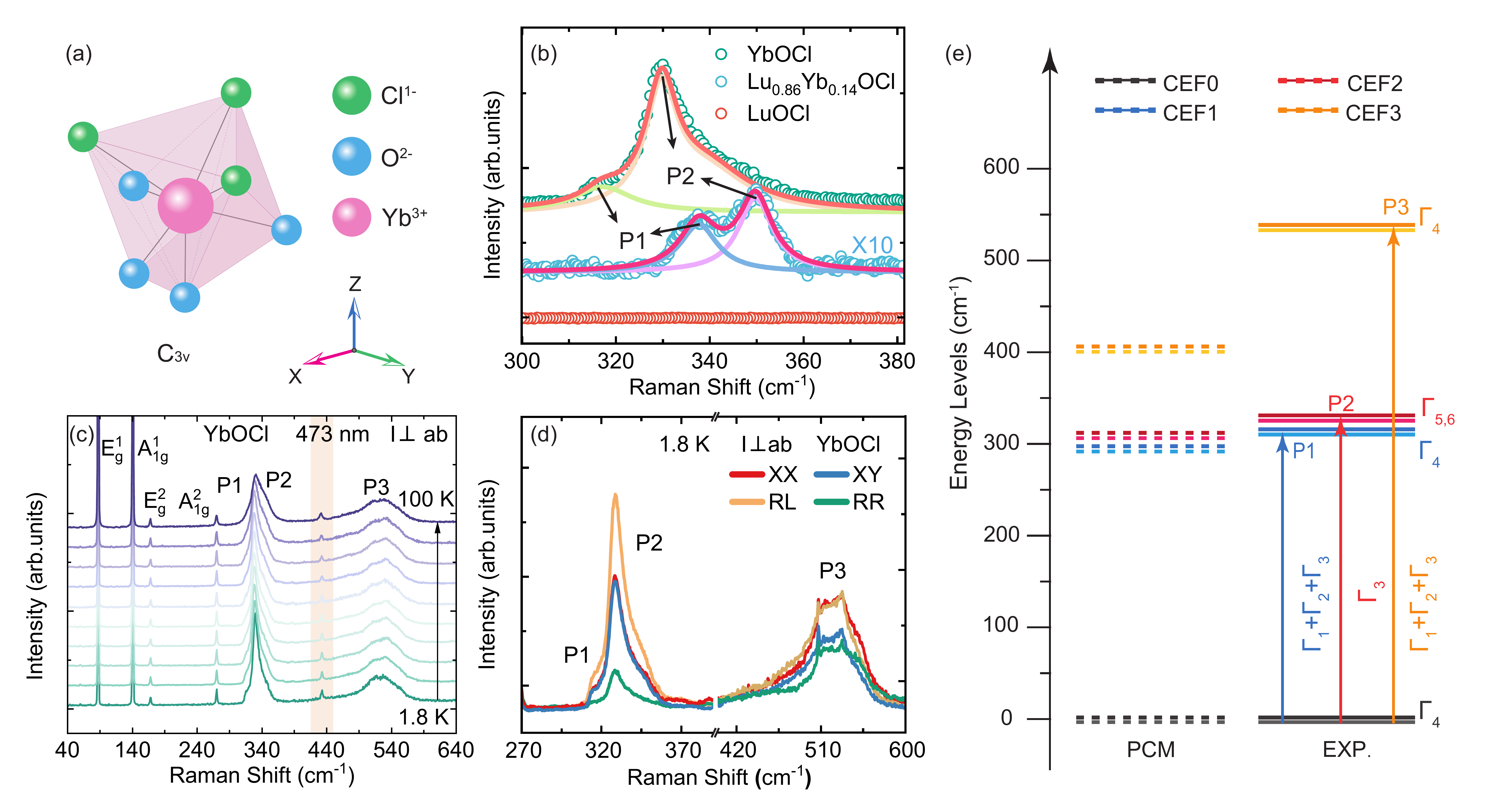}
	\caption{\label{fig:epsart}{ CEF excitations in \ce{YbOCl}. (a) The central magnetic ion \ce{Yb^{3+}} forms a CEF environment with the surrounding coordinating anions. (b) The Raman scattering peaks of the CEF first excitation energy level ($P1$) and second excitation energy level ($P2$) in \ce{YbOCl} and \ce{Lu_{0.86}Yb_{0.14}OCl}.  There are no CEF excitation peaks in \ce{LuOCl}. (c) The Raman scattering spectra of \ce{YbOCl} at different temperatures. The three CEF excitation peaks $P1$, $P2$, and $P3$ hardly change with temperature. (d) The polarization Raman spectra of \ce{YbOCl} at 1.8 K. XX and XY represent parallel and cross polarization configurations, respectively. RR and RL represent cross-circular and parallel-circular polarization configurations, respectively. (e) The CEF excitation energy levels of \ce{YbOCl}. The dashed lines are the calculated energy levels from the PCM model, while the solid lines are based on Raman scattering experimental measurements. The irreducible representations corresponding to the four CEF energy levels are $\Gamma_{4}$, $\Gamma_{4}$, $\Gamma_{5,6}$, and $\Gamma_{4}$.}}
	\label{fig:Fig3}
\end{figure*}

\subsection{CEF excitations in \ce{YbOCl}}
The electron configuration of the 4$f$ orbital in \ce{Yb^{3+}} is 4$f^{13}$.
4$f^{13}$ configuration emerges as a spectral multiplet $^{2}F_{7/2}$ with an eightfold degeneracy and another spectral multiplet $^{2}F_{5/2}$ with a xisfold degeneracy after SOC.
The energy gap between the two multiplets is about 1 eV \cite{PhysRevB.100.174436}, which means that we only need to consider the contribution associated with the multiplet $^{2}F_{7/2}$.
In \ce{YbOCl}, the central \ce{Yb^{3+}} ion and the surrounding three-coordinated \ce{Cl^{-}} anions and four-coordinated \ce{O^{2-}} anions form a local CEF environment with $C_{3v}$ point group symmetry, as shown in Fig. \hyperref[fig:Fig3]{3(a)}.
In this CEF environment, \ce{Yb^{3+}} ions with an odd number of 4$f$ electrons split into four pairs of doubly degenerate Kramers states, protected by time-reversal symmetry. 
Therefore the Hamiltonian used to describe the CEF excitations in \ce{YbOCl} is represented as follows \cite{HUTCHINGS1964227}:
\begin{equation}
	\hat{H}_{CEF} = \sum_{i} B_{2}^{0} \hat{O}_{2}^{0} + B_{4}^{0} \hat{O}_{4}^{0} + B_{4}^{3} \hat{O}_{4}^{3} + B_{6}^{0} \hat{O}_{6}^{0} + B_{6}^{3} \hat{O}_{6}^{3} + B_{6}^{6} \hat{O}_{6}^{6}
	\label{for:2}
\end{equation}
where $B_{m}^{n}$ denotes the CEF parameters and $\hat{O}_{m}^{n}$ symbolizes the Stevens operator, which is constructed based on the angular momentum $\hat{J}$ after SOC. 
In the CEF Hamiltonian, the \(z\) axis is aligned with the crystal's \(z\) axis, while the \(x\) axis is defined along the $\delta_{1}$-bond, which preserves the $C_{2}$ rotational symmetry\cite{SM}.

According to the Hamiltonian describing the CEF of \ce{YbOCl}, three CEF excitation energy levels can be detected.
This is consistent with the three excitation peaks observed in the Raman scattering spectrum of \ce{YbOCl} at 319, 327, and 523 cm$^{-1}$.
In addition, we can also demonstrate from many aspects that these three excitation peaks are CEF excitations of \ce{YbOCl}.

(i) We can perform a qualitative analysis based on the energy scale.
\ce{YbOCl} is a good insulator, and within the energy range of several hundred wavenumbers, the excitations observed in the Raman scattering spectrum can only be attributed to lattice vibrations, magnetic excitations, or CEF excitations.
Through the above analysis of lattice vibrations, we have ruled out the possibility that these three additional excitations are phonon peaks.
Moreover, the phase transition temperature of \ce{YbOCl} is 1.3 K, and the magnetic excitation measured by INS does not exceed 1 meV \cite{YbOCl}.
Therefore the excitation peaks of $P1$, $P2$, and $P3$  cannot be caused by magnetic excitations.
The CEF excitations for \ce{Yb^{3+}} ions are generally in the range of tens of millielectron volts, such as \ce{YbMgGaO4} \cite{PhysRevLett.118.107202}, \ce{NaYbS2} \cite{PhysRevB.98.220409, NaYbS2}, and \ce{NaYbSe2} \cite{PhysRevB.103.035144}. 
From the energy perspective, these three excitation peaks most closely match the characteristics of CEF excitations.
We can also rule out the possibility that these three excitation peaks are caused by fluorescence excitation, which depends on different laser wavelengths. As shown in the inset of Fig. \hyperref[fig:Fig2]{2(a)}, these three extra excitation peaks can still be observed under lasers with different wavelengths \cite{PhysRevLett.56.232}.

(ii) We also conducted temperature-dependent Raman scattering measurements on \ce{YbOCl}, as shown in Fig. \hyperref[fig:Fig3]{3(c)}. The three excitation peaks $P1$, $P2$, and $P3$ show almost no change with temperature \cite{SM}.
For example, in geometrically frustrated rare-earth titanate pyrochlores \cite{PhysRevB.77.214310}, the CEF excitations exhibit almost no change with temperature.
This is also a key piece of evidence for determining the CEF excitations in rare-earth magnets.

(iii) We calculated the CEF excitation energy levels of \ce{YbOCl} based on the PCM\cite{osti_1770657}, as indicated by the dotted lines in Fig. \hyperref[fig:Fig3]{3(e)}.
The calculation results show that the CEF first and second excitation energy levels of \ce{YbOCl} are  close to each other. The CEF third excitation energy level is far away from the first and second energy levels.
This feature is consistent with the results of our Raman scattering measurements.
The close proximity of the CEF first and second energy levels is uncommon in Yb-based rare-earth magnets, which may be attributed to the irregular local structure of \ce{YbO_4Cl_3} formed by the central \ce{Yb^{3+}} ion and surrounded by three coordinating ligand anions \ce{Cl^-} and four coordinating ligand \ce{O^{2-}},  which conforms to $C_{3v}$ point group symmetry.
Compared to rare-earth chalcogenide compounds \cite{Liu_2018}, the CEF environment of \ce{YbOCl} has a lower point group symmetry and lacks inversion symmetry.
It should be noted that the PCM is a rough approximation. 
The analysis of CEF should be based on experimental measurements.

(iv) In the diluted magnetic sample \ce{Lu_{0.86}Yb_{0.14}OCl}, we also observed weak excitation peaks near the CEF first and second energy levels of \ce{YbOCl}, as shown in Fig. \hyperref[fig:Fig3]{3(b)}. The scattering intensity in \ce{Lu_{0.86}Yb_{0.14}OCl} excitation peaks is about 10 times weaker than in \ce{YbOCl}.
This is consistent with the fundamental principle that the excitation intensity decreases as the proportion of magnetic ions is reduced.
It should be pointed out that  the excitation peak positions of the CEF first and second energy levels observed in \ce{YbOCl} are slightly different from those for \ce{Lu_{0.86}Yb_{0.14}OCl}. This is mainly related to the slight changes in lattice parameters and atom positions of the doped samples.
It can also be proved that the phonon peaks of $E_{g}^{1}$, $A_{1g}^{1}$, $E_{g}^{2}$, and $A_{1g}^{2}$ of \ce{YbOCl} and \ce{Lu_{0.86}Yb_{0.14}OCl} are not completely consistent [see Fig. \hyperref[fig:Fig2]{2(a)}].

\begin{table}[htbp]
	\centering
	\caption{The Raman tensors for CEF transitions}
	\label{S:S4}
	\renewcommand{\arraystretch}{1.5}
	\scalebox{1}{
		\begin{tabular}{ccc}
			\hline
			\hline
			\(\Gamma_1\) & \(\Gamma_2\) & \(\Gamma_3\) \\
			\hline
			\(
			\begin{pmatrix}
				a & 0 & 0 \\
				0 & a & 0 \\
				0 & 0 & b
			\end{pmatrix}
			\)
			&
			\(
			\begin{pmatrix}
				0 & c & 0 \\
				-c & 0 & 0 \\
				0 & 0 & 0
			\end{pmatrix} 
			\) 
			&
			\begin{tabular}{@{}c@{}}
				\(
				\begin{pmatrix}
					0 & d & f \\
					d & 0 & 0 \\
					f & 0 & 0
				\end{pmatrix}
				\) 
				\(
				\begin{pmatrix}
					0 & d & -f \\
					d & 0 & 0 \\
					f & 0 & 0
				\end{pmatrix}
				\) \\
				\(
				\begin{pmatrix}
					d & 0 & 0 \\
					0 & -d & f \\
					0 & f & 0
				\end{pmatrix}
				\) 
				\(
				\begin{pmatrix}
					d & 0 & 0 \\
					0 & -d & f \\
					0 & -f & 0
				\end{pmatrix}
				\)\\
				(symmetric) (antisymmetric)
			\end{tabular} \\
			\hline
			\hline
		\end{tabular}
	}
	\label{table:TABLE I}
\end{table}

\begin{table*}
	\centering
	\caption{CEF parameters of \ce{YbOCl} (in meV) and Stevens normalization (in \ce{cm^{-1}})\cite{SM}}
	\renewcommand{\arraystretch}{1.5}
	\begin{tabularx}{\textwidth}{ZZZZZZZ}
		\hline
		\hline
		unit & \( B_2^0 \)  & \( B_4^0 \)  & \( B_4^3 \)  & \( B_6^0 \)  & \( B_6^3 \)  & \( B_6^6 \)  \\
		\hline
		meV & $-3.163 \times 10^{-1}$ & $-1.900 \times 10^{-3}$ & $-7.068 \times 10^{-1}$ & $ 1.000 \times 10^{-4}$ & $-5.200 \times 10^{-3}$ & $6.300 \times 10^{-3}$ \\
		\hline
		unit & \( A_2^0 \langle r^2 \rangle \)  & \(  A_4^0 \langle r^4 \rangle \)  & \(  A_4^3 \langle r^4 \rangle \)  & \(  A_6^0 \langle r^6 \rangle \)  & \(  A_6^3 \langle r^6 \rangle \)  & \(  A_6^6 \langle r^6 \rangle \)  \\
		 & = \( B_2^0 \)/$\alpha$\textcolor{blue}{$^a$}  & = \( B_4^0 \)/$\beta$\textcolor{blue}{$^b$}  & = \( B_4^3 \)/$\beta$  & = \( B_6^0 \)/$\gamma$\textcolor{blue}{$^c$}  & = \( B_6^3 \)/$\gamma$  & = \( B_6^6 \)/$\gamma$  \\
		\hline
		cm$^{-1}$ &$-80.372$ & $8.850$ & $3292.274$ & $ 5.450 $ & $-283.400$ & $343.350$  \\
		\hline
		\hline
	\end{tabularx}
	\footnotetext[1]{$1/\alpha$ = $2.541 \times 10^2$}
	\footnotetext[2]{$1/\beta$ = $-4.658 \times 10^3$}
	\footnotetext[3]{$1/\gamma$ = $5.450 \times 10^4$}
	\label{table:TABLE II}
\end{table*}

\begin{table*}
	\centering
	\caption{ CEF energy levels, wave functions, and symmetry of \ce{YbOCl}}
	\renewcommand{\arraystretch}{1.5}
	\begin{tabularx}{\textwidth}{ZcYc} 
		\hline
		\hline
		& \textbf{Energy (cm$^{-1}$)} & \textbf{Wavefunction} & \textbf{Symmetry} \\
		\hline
		CEF0 & 0 (0 meV) & $ \left| \psi_{0, \pm} \right\rangle = \pm0.6738 \left| \pm\frac{7}{2} \right\rangle + 0.6274 \left| \pm\frac{1}{2} \right\rangle \mp 0.3905 \left| \mp\frac{5}{2} \right\rangle$ & $\Gamma_4$ \\
		\multirow{2}{*}{CEF1} & \multirow{2}{*}{318.63 (39.51 meV)} & $\left| \psi_{1, \pm} \right\rangle = 0.4882 \left| \pm\frac{7}{2} \right\rangle \mp 0.2291 \left| \pm\frac{5}{2} \right\rangle \mp 0.0070 \left| \pm\frac{1}{2}  \right\rangle - 0.0019 \left| \mp\frac{1}{2} \right\rangle + 0.8312 \left| \mp\frac{5}{2} \right\rangle \mp 0.1346 \left| \mp\frac{7}{2} \right\rangle $ & \multirow{2}{*}{$\Gamma_4$} \\
		CEF2 & 326.94 (40.54 meV) & $ \left| \psi_{2, \pm} \right\rangle= \pm0.9745 \left| \pm\frac{3}{2} \right\rangle - 0.2245 \left| \mp\frac{3}{2} \right\rangle$ & $\Gamma_{5,6}$ \\
		CEF3 & 522.73 (64.76 meV) & $ \left| \psi_{3, \pm} \right\rangle= \pm 0.5381 \left| \pm\frac{7}{2} \right\rangle - 0.7787 \left| \pm\frac{1}{2} \right\rangle \mp 0.3226 \left| \mp\frac{5}{2} \right\rangle $ & $\Gamma_4$ \\
		\hline
		\hline
	\end{tabularx}
	\label{table:TABLE III}
\end{table*}
(V) More importantly, we can proceed from the CEF Raman tensor for further analysis.
The irreducible representations of the four CEF excitation levels of \ce{YbOCl} are $\Gamma_{4}$, $\Gamma_{4}$, $\Gamma_{5,6}$, and $\Gamma_{4}$, respectively.
According to the Raman transition rules \cite{eyring2002handbook, schaack2007raman}:
\begin{equation}
	\Gamma_{i} \otimes \Gamma_{f} \subseteq \Gamma_{Raman}.
	\label{for:3}
\end{equation}
The transitions from $\Gamma_{4}$ to $\Gamma_{4}$ and $\Gamma_{4}$ to $\Gamma_{5,6}$ can be decomposed into the following irreducible representation \cite{Koningstein1972IntroductionTT, becker1986electronic}:
\begin{equation}
	\begin{aligned}
		&\Gamma_4 \rightarrow \Gamma_4 = \Gamma_1 \oplus \Gamma_2 \oplus \Gamma_3, \\
		&\Gamma_4 \rightarrow \Gamma_{5,6} = \Gamma_3.
	\end{aligned}
	\label{for:4}
\end{equation}
The corresponding Raman tensors of $\Gamma_1$, $\Gamma_2$, and $\Gamma_3$ are shown in Table \ref{table:TABLE I}.

Among them, $\Gamma_3$ possesses two types of Raman tensors: symmetric and antisymmetric \cite{schaack2007raman, Koningstein1972IntroductionTT}.
Furthermore, according to these Raman tensors, we can determine the Raman scattering intensities for XX, XY, cross-circular (RL), and parallel-circular (RR) polarizations. For the Raman transition between $\Gamma_{4}$ and $\Gamma_{4}$ states:
\begin{equation}
	\begin{aligned}
		I^{XX}_{\text{$\Gamma_4 \rightarrow \Gamma_4$}} & = a^2 + 4d^2  & I^{XY}_{\text{$\Gamma_4 \rightarrow \Gamma_4$}} & = 4d^2 + c^2,  \\
		I^{RR}_{\text{$\Gamma_4 \rightarrow \Gamma_4$}} & = a^2 + c^2  &
		I^{RL}_{\text{$\Gamma_4 \rightarrow \Gamma_4$}} & = 8d^2.  \\ 
	\end{aligned}
	\label{for:5}
\end{equation}

\noindent For the Raman transition between $\Gamma_{4}$ and $\Gamma_{5,6}$ states:

\begin{equation}
	\begin{aligned}
        I^{XX}_{\text{$\Gamma_4 \rightarrow \Gamma_{5,6}$}} &= 4d^2  &
		I^{XY}_{\text{$\Gamma_4 \rightarrow \Gamma_{5,6}$}} &= 4d^2,  \\
		I^{RR}_{\text{$\Gamma_4 \rightarrow \Gamma_{5,6}$}} &= 0  &
		I^{RL}_{\text{$\Gamma_4 \rightarrow \Gamma_{5,6}$}} &= 8d^2.  &
	\end{aligned}
	\label{for:6}
\end{equation}
The $P1$ Raman scattering peak, corresponding to the CEF first excitation energy level, follows the transition rule from $\Gamma_{4}$ to $\Gamma_{4}$. According to the transition from $\Gamma_{4}$ to $\Gamma_{4}$ described in formula \hyperref[for:5]{(5)}, the sum of scattering intensities for XX and XY polarization is equal to the sum of scattering intensities for RR and RL polarization.
In Fig. \hyperref[fig:Fig3]{3(d)}, the sum of the scattering intensities of the $P1$ peak under XX and XY polarizations is basically equal to the sum of the scattering intensities under RL and RR polarizations.
Similarly, the $P3$ Raman scattering peak, which represents the CEF third excitation energy level, also follows the transition rule from $\Gamma_{4}$ to $\Gamma_{4}$. 
The sum of the scattering intensities of the $P3$ peak under XX and XY polarizations is also consistent with the sum of the scattering intensities under RL and RR polarizations.
For the $P2$ scattering peak, which corresponds to the CEF second excitation energy level, it conforms to the transition rule from $\Gamma_{4}$ to $\Gamma_{5,6}$.
Under XX and XY polarizations, the scattering intensity for the transition from $\Gamma_{4}$ to $\Gamma_{5,6}$ remains unchanged. The scattering intensity under RL polarization is twice that of XX and XY polarizations, while under RR polarization, the scattering intensity is zero.
The scattering intensity of the $P2$ peak under different polarizations basically conforms to this characteristic. It should be noted that the $P2$ peak does not completely extinguish in the RR mode, which is related to our Raman spectrometer and optical path. Similarly, the phonon peaks under different polarizations in Fig. \hyperref[fig:Fig2]{2} also exhibit a similar issue.

Through the above analysis, we have demonstrated from multiple perspectives that $P1$ ($\sim$ 319 cm$^{-1}$), $P2$ ($\sim$ 327 cm$^{-1}$), and $P3$ ($\sim$ 523 cm$^{-1}$) Raman scattering peaks correspond to the CEF excitations of \ce{YbOCl}.
Furthermore, since the CEF Raman tensor differs from that of phonons, our analysis based on the CEF Raman tensor provides strong evidence for identifying CEF excitations.

Based on the three CEF energy levels determined by Raman scattering, we can fit the $B_{m}^{n}$ parameters in the CEF Hamiltonian of \ce{YbOCl} and calculate the corresponding CEF wave functions. 
The CEF parameters, energy levels, wave functions, and irreducible representations of \ce{YbOCl} are presented in Table \ref{table:TABLE II} and \ref{table:TABLE III}.
For convenience, the CEF parameters, divided by the corresponding reduced matrix elements (Stevens normalization), are also present in Table \ref{table:TABLE II}\cite{SM}.
To verify the reliability of these fitted CEF parameters, we also calculated the \(g\) factors along the \(c\) axis and in the \(ab\) plane based on the computed CEF ground state wave functions.
The relationship between the \(g\) factors and CEF ground state wave functions is as follows \cite{PhysRevB.103.184419}:
\begin{flalign}
	&& g_{c} &= 2g_{j} \left| \langle \psi_{1,\pm} | J_{z} | \psi_{1,\pm} \rangle \right| ,&\nonumber \\
	&& g_{\text{ab}} &= g_{j} \left| \langle \psi_{1,\pm} | J_{\pm} | \psi_{1,\mp} \rangle \right|, &
	\label{for:7}
\end{flalign}
where $g_{j} = 8/7$ and $ \left| {\left. {{\psi _{1, \pm }}} \right\rangle } \right.\ $ are the Landé \(g\) factor for free \ce{Yb^{3+}} and the CEF ground state wave functions.
We calculated the \(g\) factors along the \(c\) axis and in the \(ab\) plane to be $g_{c} \sim 3.2$ and $g_{ab} \sim 3.4$, respectively.
This is in agreement with the results of the \(g\) factors along the \(c\) axis ($g_{c}$ $\sim$ 3.2) and in the \(ab\) plane ($g_{ab}$ $\sim$ 3.5) from ESR measurements on the diluted magnetic sample \ce{Lu_{0.86}Yb_{0.14}OCl} at 1.8 K \cite{SM}.
Interestingly, we also observed a series of satellite peaks near the main resonance peaks in \ce{Lu_{0.86}Yb_{0.14}OCl}, which may be associated with pairs formed by the magnetic ions \ce{Yb^{3+}}\cite{SM}.
This further reinforces the reliability of the $B_{m}^{n}$ parameters in the CEF Hamiltonian obtained from our fitting, and also lays the foundation for analysis of the splitting of the CEF energy levels of \ce{YbOCl} under the magnetic fields.
\begin{figure}[h]
	\includegraphics[scale=0.38]{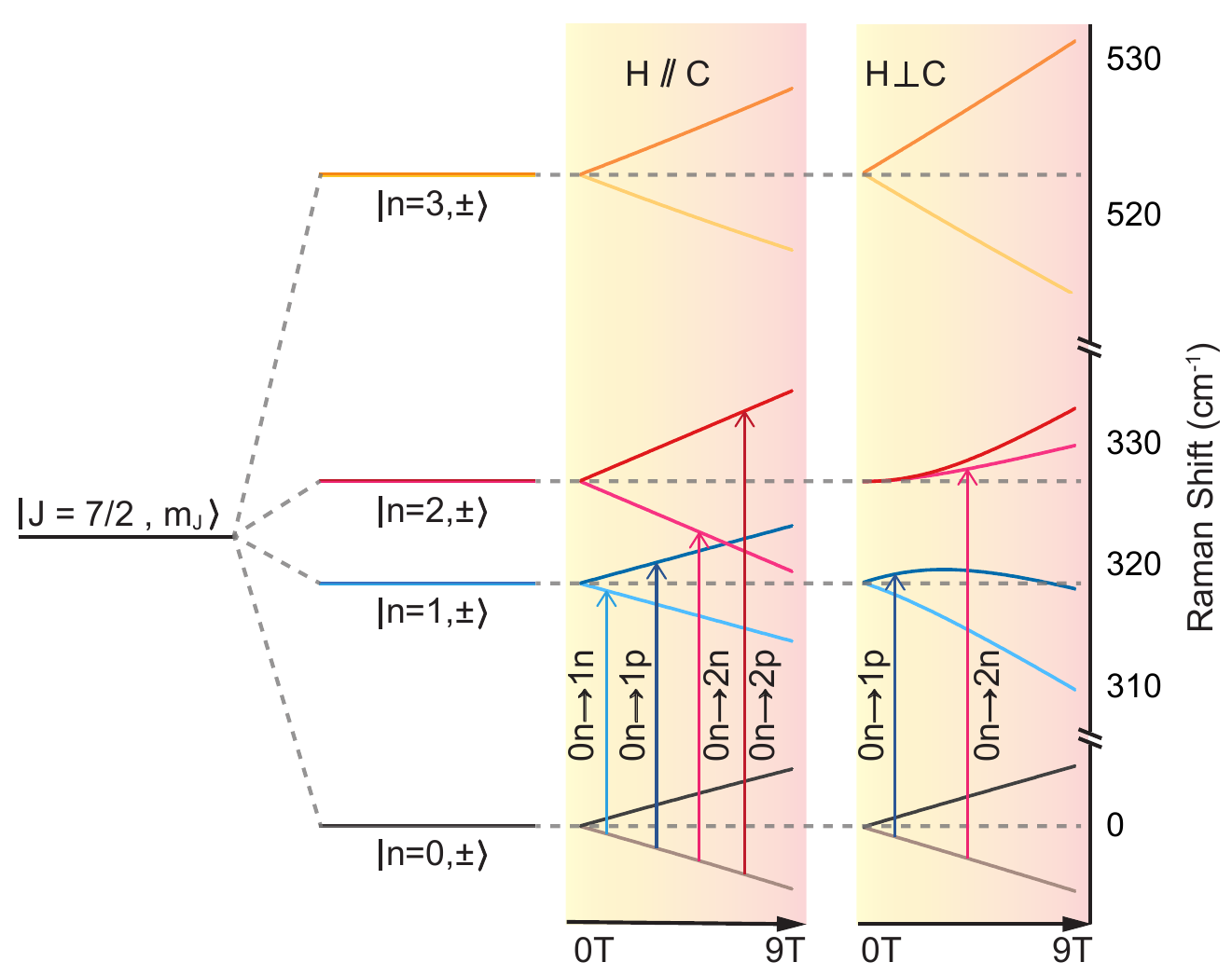}
	\caption{\label{fig:epsart}{Schematic diagram of the CEF energy spectrum of \ce{Yb^{3+}}. The CEF local environment of \ce{Yb^{3+}} with $C_{3v}$ point group symmetry splits the \ce{Yb^{3+}} ground-state manifold into four Kramers doublets. The doubly degenerate CEF energy levels split linearly under a magnetic field applied along the \(c\) axis, while they show nonlinear splitting under a magnetic field in the \(ab\) plane. The arrows represent transitions between different CEF levels.}}
	\label{fig:Fig4}
\end{figure}

\begin{figure*}[t]
	\includegraphics[scale=0.35]{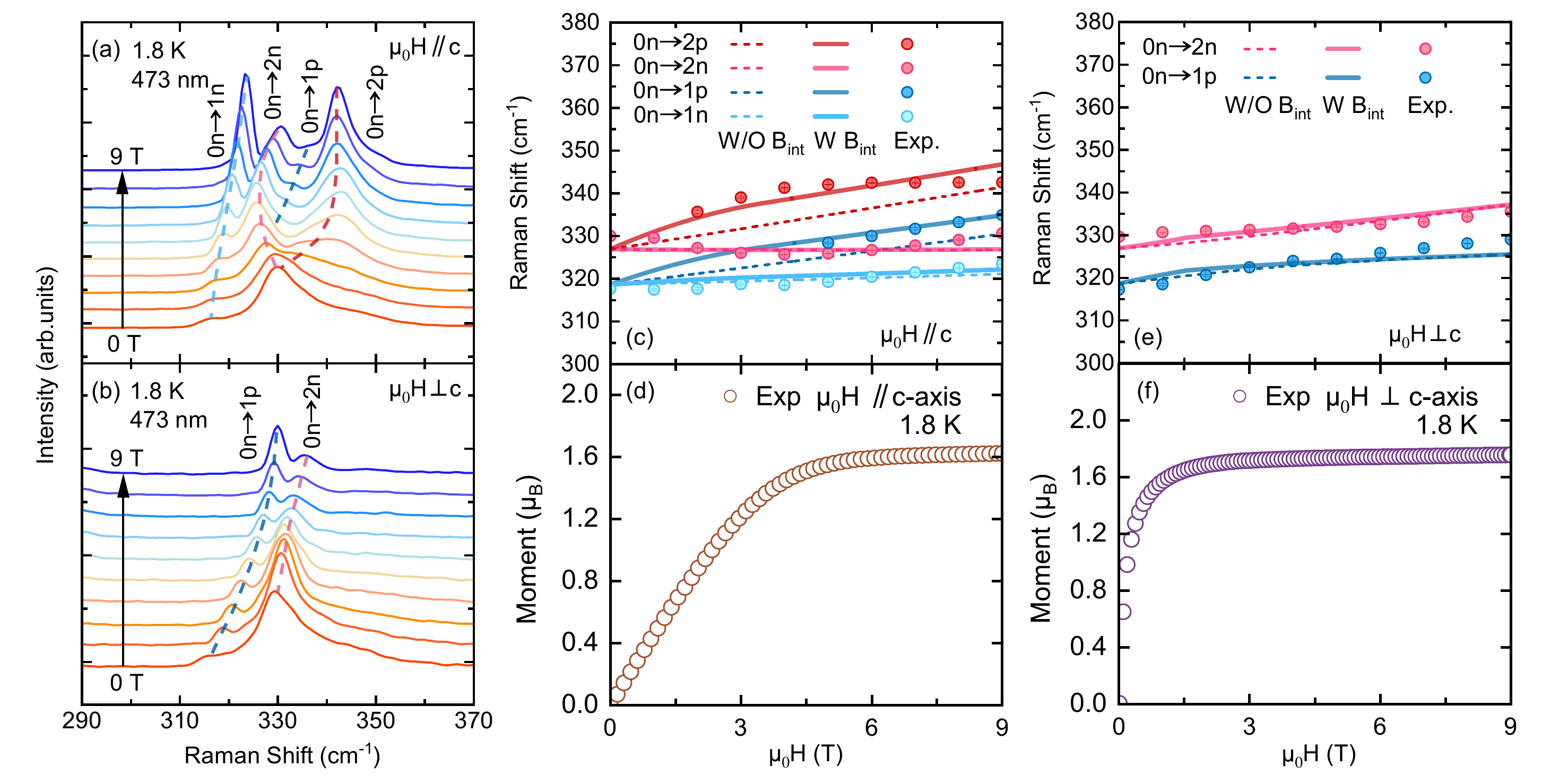}
	\caption{\label{fig:epsart}{Magnetic field-induced splitting of CEF energy levels in \ce{YbOCl} and M-H data. [(a) and (b)] Raman scattering spectra of YbOCl under magnetic fields applied along the \(c\) axis and in the \(ab\) plane at 1.8 K, respectively. [(c) and (e)] The central peak positions of the excitation peaks change with magnetic fields along the \(c\) axis and in the \(ab\) plane, respectively. The circles are experimental data. The dashed lines represent the calculated results considering only the external magnetic fields. The solid lines are the calculated results considering both the external and internal magnetic fields. [(d) and (f)] M-H data of \ce{YbOCl} along the \(c\) axis and in the \(ab\) plane, respectively.}}
	\label{fig:Fig5}
\end{figure*}
For the CEF wave functions we calculated in Table \ref{table:TABLE III}, we can do more discussion.
The irreducible representations of the wave functions of the CEF ground state, first excited state, and third excited state are all $\Gamma_{4}$.
The irreducible representation of the wavefunction of the CEF second excited state is $\Gamma_{5,6}$.
The CEF wavefunction with the $\Gamma_{4}$ irreducible representation contains the three components $\left| {\left. { \pm \frac{7}{2}} \right\rangle } \right.$, $\left| {\left. { \pm \frac{5}{2}} \right\rangle } \right.$, and $\left| {\left. { \pm \frac{1}{2}} \right\rangle } \right.$.
The CEF wave functions with the $\Gamma_{5,6}$ irreducible representation only contain the $\left| { \pm \frac{3}{2}} \right\rangle$  component \cite{PhysRevLett.112.167203, doi:10.1080/00018730500037264, PhysRevLett.122.187201, PhysRevLett.115.097202}.
Due to the inherent complexity of the CEF wave functions, the distribution of charge density in space exhibits strong anisotropy \cite{PhysRevB.100.180406}.
Therefore these degenerate CEF energy levels may split in a nonl-inear way under the application of external magnetic fields.
As shown in Fig. \hyperref[fig:Fig4]{4}, we calculated the splitting of the CEF energy levels of \ce{YbOCl} under magnetic fields applied along the \(c\) axis and in the \(ab\) plane.
From the figure, we can clearly see that the doubly degenerate CEF energy levels split in a linear manner under magnetic fields along the \(c\) axis. 
However, the splitting of these levels exhibits significant nonlinear effects under magnetic fields in the \(ab\) plane.
From the perspective of CEF wave functions, this phenomenon can be easily understood.
$\left| J = 7/2, m_{j} \right\rangle$ is an eigenstate of $\hat{J}_{z}$ operator. In quantum mechanics, the state $\left| J = 7/2, m_{j} \right\rangle$ represents the total angular momentum $J$ and its projection $m_{j}$ along the \(c\) axis.
Here, $\hat{J}_{z}$ is the operator corresponding to the z-component of the angular momentum.
For an eigenstate of $\hat{J}_{z}$, we have:
\begin{equation}
	\hat{J}_{z} \left| J = 7/2, m_{j} \right\rangle = m_{j} \hbar \left| J = 7/2, m_{j} \right\rangle.
	\label{for:8}
\end{equation}
This equation indicates that $\left| J = 7/2, m_{j} \right\rangle$ is indeed an eigenstate of $\hat{J}_{z}$ with eigenvalue $m_{j}\hbar$.
Therefore applying a magnetic field along the \(c\) axis will result in linear splitting of these doubly degenerate CEF energy levels.
However, $\left| J = 7/2, m_{j} \right\rangle$ is not an eigenstate of the $\hat{J}_{x}$ or $\hat{J}_{y}$ operators, meaning that these states will undergo nonlinear splitting when magnetic fields are applied in the \(ab\) plane, which involves components of the angular momentum in the \(ab\) plane \cite{PhysRevResearch.3.043202}.
It is worth noting that about the 6 T magnetic field causes the split CEF first and second excitation levels to cross along the \(c\) axis.
This is closely related to the proximity of the CEF first and second excitation levels, and we also observed this phenomenon in Raman scattering under magnetic fields.
Next, we will carry out further research on the splitting behavior of these CEF energy levels by combining Raman scattering experiments under magnetic fields.

\subsection{Magnetic field-induced nonlinear splitting of CEF energy levels}
Through Raman scattering experiments, we have identified the three CEF excitation energy levels of \ce{YbOCl}.
Due to the inherent anisotropy of the CEF wave functions, the nonlinear splitting characteristics exhibited under magnetic fields are worthy of further investigation.
Raman scattering is almost the only method to study the CEF splitting under magnetic fields.
In general, the splitting of the CEF under the magnetic field is about a dozen wavenumbers (1 $\sim$ 2 meV), which is comparable to the energy resolution of high-energy INS. 
The energy resolution of Raman scattering is much better than that of INS, so the weak shift in energy level can be detected by Raman scattering.
Besides, Raman scattering is more convenient for studying the splitting of CEF energy levels under magnetic fields compared with INS.
Therefore we performed Raman scattering measurements under different magnetic fields along the \(c\) axis and in the \(ab\) plane of \ce{YbOCl} at 1.8 K, as shown in Figs. \hyperref[fig:Fig5]{5(a)} and \hyperref[fig:Fig5]{5(b)}.
For the Raman scattering spectra along the \(c\) axis under different magnetic fields, we observed that the original two CEF excitations, $P1$ and $P2$, split into four excited peaks with the increase of the magnetic fields.
The four emerging excitation peaks under the magnetic fields are labeled as $0n$ $\rightarrow$ $1n$ ($\mu_0 H \parallel c$), $0n$ $\rightarrow$ $2n$ ($\mu_0 H \parallel c$), $0n$ $\rightarrow$ $1p$ ($\mu_0 H \parallel c$), and $0n$ $\rightarrow$ $2p$ ($\mu_0 H \parallel c$).
Among them, $0n$ $\rightarrow$ $1n$ ($\mu_0 H \parallel  c$) and $0n$ $\rightarrow$ $1p$ ($\mu_0 H \parallel c$) shift almost linearly towards higher energies with increasing magnetic field.
In particular, the peaks of $0n$ $\rightarrow$ $2n$ ($\mu_0 H \parallel c$) and $0n$ $\rightarrow$ $2p$ ($\mu_0 H \parallel c$) show significant nonlinear behaviors as the external magnetic field increases.
For the Raman spectra with the external magnetic field in the \(ab\) plane, both the excitation peaks of $0n$ $\rightarrow$ $1p$ ($\mu_0 H \perp c$) and $0n$ $\rightarrow$ $2n$ ($\mu_0 H \perp c$) shift towards higher energies with the increase of the magnetic field.

The calculated results, which only account for the CEF splitting effects under the external magnetic fields, are depicted as dashed lines in Figs. \hyperref[fig:Fig5]{5(c)} and \hyperref[fig:Fig5]{5(e)}. 
It can be seen that for the two excitations, $0n$ $\rightarrow$ $1p$ ($\mu_0 H \parallel c$) and $0n$ $\rightarrow$ $2p$ ($\mu_0 H \parallel c$), the calculated results represented by the dashed lines show a clear deviation from the experimental measurements.
This observation prompts us to further investigate the potential reasons for the deviations between the computational predictions and experimental measurements.
A reasonable interpretation is that the discrepancies arise from internal magnetic fields contributed by intrinsic magnetic moments.
To validate this idea, we measured the M-H data along the \(c\) axis and in the \(ab\) plane of \ce{YbOCl} at 1.8 K, as shown in Figs. \hyperref[fig:Fig5]{5(d)} and \hyperref[fig:Fig5]{5(f)}. 
In both  the \(c\) axis and \(ab\) plane, the magnetic moments exhibit an initial increase with the applied magnetic field followed by gradual saturation. 
Saturation is reached at approximately 5 T along the \(c\) axis and 2 T in the \(ab\) plane.
The saturation magnetic moments along the \(c\) axis and in the \(ab\) plane are approximately 1.6 $\mu_B$ and 1.7 $\mu_B$, respectively.
Therefore it is necessary to consider the influence of the internal magnetic fields contributed by intrinsic magnetic moments on the splitting of the CEF.

To account for the effects of external and internal magnetic fields on the CEF, we selected the XXZ model with nearest neighbor spin interactions.
The Hamiltonian of the XXZ model and the Zeeman term have the following form \cite{PhysRevB.100.134434}:
\begin{widetext}
	\begin{equation}
		\begin{array}{l}
			{\hat H_{XXZ}} + {\hat H_{Zeeman}} = \sum\limits_{\left\langle {ij} \right\rangle } {{\mathcal{J}_ \pm }\left( {\hat J_i^ + \hat J_j^ -  + \hat J_i^ - \hat J_j^ + } \right) + {\mathcal{J}_{zz}}\hat J_i^z\hat J_j^z}  - {\mu _0}{\mu _B}{g_j}\sum\limits_i {{h_x}\hat J_i^x + {h_y}\hat J_i^y + {h_z}\hat J_i^z},
		\end{array}
	\end{equation}
	\label{for:9}
\end{widetext}
where $\mathcal{J}_\pm$ and $\mathcal{J}_{zz}$ represent the nearest neighbor spin-exchange interaction with angular momentum $J = 7/2$ in the \(ab\) plane and along the \(c\) axis, respectively.
$h_x$, $h_y$, and $h_z$ are external magnetic fields applied along the spin $\hat{J}_{x}$, $\hat{J}_{y}$, and $\hat{J}_{z}$ directions, respectively.
From the perspective of MF theory, we can further simplify the Hamiltonian described above.
The MF Hamiltonian along the \(c\) axis and in the \(ab\) plane is expressed as follows:
\begin{equation}
	\begin{array}{l}
		{{\hat H}_{c,MF}} = {{\hat H}_{CEF}} - {\mu _0}{\mu _B}{g_j}\left( {{h_z} + {h_{eff,c}}} \right)\sum\limits_i {\hat J_i^z} \\
		\\
		{{\hat H}_{ab,MF}} = {{\hat H}_{CEF}} - {\mu _0}{\mu _B}{g_j}\left( {{h_x} + {h_{eff,ab}}} \right)\sum\limits_i {\hat J_i^x} 
	\end{array}
	\label{for:10}
\end{equation}
where $h_{eff,c} = \lambda_{c} M_{c}$ and $h_{eff,ab} = \lambda_{ab} M_{ab}$ are the effective internal magnetic fields along the \(c\) axis and in the \(ab\) plane, respectively\cite{SM}.  $\lambda_{c}$ and $\lambda_{ab}$ are effective field coefficients related to spin-exchange interaction. $M_{c}$ and $M_{ab}$ are the magnetic moments along the \(c\) axis and in the \(ab\) plane, respectively. 
Based on the MF Hamiltonian described in Formula \hyperref[for:10]{(10)}, we applied diagonalization techniques to refit the experimental Raman scattering data (circles) presented in Fig. \hyperref[fig:Fig5]{5(c)} and \hyperref[fig:Fig5]{5(e)}, yielding the simulation results represented by solid lines.
After considering the internal magnetic field, the simulation results more closely align with the experimental data. 
Especially with the magnetic field applied along the \(c\) axis, the calculations considering the internal magnetic field show significant improvement for the $0n$ $\rightarrow$ $1p$ ($\mu_0 H \parallel c$) and $0n$ $\rightarrow$ $2p$ ($\mu_0 H \parallel c$) excitations.
Compared to the magnetic field along the \(c\) axis, the corrections considering the internal magnetic field in the \(ab\) plane are not significant. The reason is that the $0n$ $\rightarrow$ $1p$ ($\mu_0 H \perp c$) and $0n$ $\rightarrow$ $2p$ ($\mu_0 H \perp c$) excitations in the \(ab\) plane themselves do not exhibit significant changes with the magnetic field.\par
Simultaneously, we estimated the effect of magnetic dipole interactions, based on the saturated magnetic moment of Yb$^{3+}$ $\sim$ 1.7 $\mu_B$. The nearest-neighbor distance between Yb$^{3+}$ ions is 3.55 \AA. Using the calculation formula (S.13) \cite{SM}, we estimate that the energy of the dipolar interactions is approximately 0.06 K, which is significantly smaller than the energy scale of spin-exchange interactions in YbOCl (approximately 1--2 K). Therefore the influence of magnetic dipole interactions can be considered negligible.
To be honest, although we have considered the influence of the internal magnetic field within the MF framework, there are still some discrepancies between the calculated results and the experimental measurements. 
In fact, for rare-earth honeycomb lattice magnets, the influence of off-diagonal spin-exchange interactions should be considered. The MF approximation clearly cannot incorporate the off-diagonal spin-exchange interactions \cite{PhysRevB.99.224409}.
However, from a physical perspective, the MF approximation provides valuable insights for understanding the CEF nonlinear splitting under magnetic fields in \ce{YbOCl}.

\subsection{Summary}

In this paper, we systematically investigate the CEF excitations and magnetic field-induced nonlinear splitting of CEF excitations in \ce{YbOCl} using Raman scattering techniques.
Firstly, through the temperature-dependent XRD patterns of \ce{YbOCl}, we can rule out temperature-induced structural phase transitions.
At the same time, through low-temperature Raman scattering experiments on \ce{YbOCl}, the nonmagnetic control sample \ce{LuOCl}, and the diluted magnetic sample \ce{Lu_{0.86}Yb_{0.14}OCl}, as well as the analysis of lattice vibrations, we can identify the phonon peaks in the Raman scattering spectrum of \ce{YbOCl}.
Additionally, we observed three extra excitation peaks in the Raman scattering spectrum of \ce{YbOCl}.
Secondly, we demonstrated from multiple perspectives that these three extra excitation peaks are attributed to the CEF excitations of \ce{YbOCl}.
We identified the three CEF excitation energy levels of \ce{YbOCl} to be approximately 319, 327, and 523 cm$^{-1}$.
Moreover, we can obtain the CEF parameters and CEF wave functions of \ce{YbOCl} by combining the CEF theory and diagonalization techniques.
Furthermore, we studied the splitting behavior of the CEF excitations of \ce{YbOCl} under magnetic fields.
The CEF excitations of \ce{YbOCl} exhibit significant nonlinear splitting characteristics under magnetic fields, especially when the magnetic field is along the \(c\) axis.
From the MF perspective, considering the influence of the internal magnetic field, we can effectively explain the nonlinear splitting effect of CEF excitations in \ce{YbOCl} under magnetic fields.

Our study of CEF excitations in \ce{YbOCl} using Raman scattering techniques is not only applicable to this material but also serves as a model for investigating CEF excitations in other rare-earth materials.
This makes it possible for us to systematically study the CEF exciations of rare-earth magnetic ions and the coupling between the CEF and other physical quantities, such as CEF-phonon coupling.


\subsection{Acknowledgments}
This work was supported by the National Key Research and Development Program of China (Grant No. 2022YFA1402704), the National Natural Science Foundation of China (Grant No. 12274186), the Strategic Priority Research Program of the Chinese Academy of Sciences (Grant No. XDB33010100), and the Synergetic Extreme Condition User Facility (SECUF). The authors acknowledge the support provided by the Supercomputing Center of Lanzhou University.  

\bibliography{references.bib}

\end{document}